\documentstyle[aps,12pt,preprint,epsfig]{revtex}

\draft
\begin{document}

\title{Inelastic electroproduction of $\eta_c$ at $ep$ colliders}
\author{Li-Kun Hao and Feng Yuan}
\address{\small {\it Department of Physics, Peking University, Beijing
100871, People's Republic of China}}
\author{Kuang-Ta Chao}
\address{\small {\it China Center of Advanced Science and Technology
(World Laboratory), Beijing 100080, People's Republic of China\\
and Department of Physics, Peking University, Beijing 100871,
People's Republic of China}}
\maketitle

\begin{abstract}

Using the nonrelativistic QCD factorization formalism, we calculate the electroproduction
cross sections of $\eta_c$ in $ep$ collisions, including the contribution from both the
transverse photon and the longitudinal photon.
For this process the color-singlet contribution
vanishes up to the next to leading order perturbative QCD calculations.
The dominant contribution comes from the
color-octet ${}^1S_0^{(8)}$ subprocess.
The nonperturbative color-octet matrix element
of ${}^1S_0^{(8)}$ of $\eta_c$ is related to
that of ${}^3S_1^{(8)}$ of $J/\psi$ by heavy quark
spin symmetry, and the latter can be determined from the direct production
of $J/\psi$ at large transverse momentum at the Fermilab Tevatron.
The measurement of this process at DESY HERA can be viewed as another independent
test for the color-octet production mechanism.
\end{abstract}

\section{Introduction}

Studies of heavy quarkonium production in high energy collisions may provide
important information on both perturbative and nonperturbative QCD.
Recent progress in this area was stimulated by the experimental results of the
Collider Detector at Fermilab ({\bf CDF}) at the Fermilab Tevatron. During the 
Tevatron run I, the 
{\bf CDF} data \cite{cdf} for the prompt production rates of $J/\psi$ and
$\psi^\prime$ at large transverse momentum were observed to be orders of
magnitude larger than the lowest order perturbative calculation based on 
the color-singlet model\cite{RB}. At the same time, a new framework in treating
quarkonium production and decays has been advocated by Bodwin, Braaten,
and Lepage in the context of nonrelativistic quantum chromodynamics
(NRQCD) \cite{bbl}.
In this approach, the production process is
factorized into short and long distance parts, while the latter is
associated with the nonperturbative matrix elements of four-fermion
operators. 
So, for heavy quarkonium production, the quark-antiquark pair does not 
need to be in the color-singlet state in the short distance production stage,
which occurs at the scale of $1/m_Q$ ($m_Q$ being the heavy quark mass).
At this stage, it allows a color configuration other than the singlet 
for the heavy quark pair, that is the color-octet.
The later situation for heavy quarkonium production is called the 
color-octet mechanism.
In this production mechanism, a heavy quark-antiquark pair is produced
at short distances in a color-octet state, and then hadronizes into a 
final state quarkonium (the physical state) nonperturbatively. 
The color-octet terms in the gluon fragmentation to $J/\psi$($\psi'$) have been 
considered to explain the $J/\psi$($\psi'$) surplus problems discovered
by CDF \cite{surplus,s1}. Taking the nonperturbative matrix elements $\langle {\cal
O}^{J/\psi}_8(^3S_1)\rangle $ and $\langle {\cal O}^{\psi'}_8(^3S_1)\rangle $ as input
parameters, the CDF surplus problem for $J/\psi$ and $\psi'$ can be explained
as the contributions of color-octet terms due to gluon fragmentation.
In the past few years, applications of the NRQCD factorization formalism
to $J/\psi$($\psi^\prime$) production at various experimental facilities
have been studied.

Even though the color-octet mechanism has achieved some successes in describing
the production and decay of heavy quarkonia, more tests of this mechanism are still needed.
Recently, the photoproduction data from the DESY $ep$ collider HERA \cite{photo} posed
a question about the color-octet predictions for the inelastic photoproduction
of $J/\psi$\cite{photo2,MK}.
More recently, possible solutions for this problem have been suggested 
in\cite{explain,explain2,explain3}.
In this situation it is certainly helpful to find
other processes to test the color-octet mechanism in heavy 
quarkonium production.
Under this context, we have studied $\eta_c$ photoproduction in NRQCD\cite{photo-etac}.
We find that this process has several unique properties for the test of the
color-octet production mechanism: (1) this process is purely a color-octet
process up to the next to leading order in perturbative QCD;
(2) this process is dominated by the ${}^1S_0^{(8)}$ channel,
for which the color-octet matrix element
can be related to the $J/\psi$ hadroproduction at large $p_T$ by the heavy
quark spin symmetry;
(3) unlike the inelastic $J/\psi$ photoproduction 
which is affected by large diffraction background, the inelastic
photoproduction of $\eta_c$ has much lower diffraction background.
So, the measurement of this process may clarify the existing conflict between the
color-octet prediction and the experimental result on the $J/\psi$
photoprodution.
In this paper, we will extend our previous study of $\eta_c$ production from
the photoproduction region ($Q^2\approx 0$) to the deep inelastic scattering (DIS)
region ($Q^2>0$) at the energies relevant to the HERA collider.
The electroproduction of $\eta_c$ has the same features as the above mentioned
for the test of the color-octet production mechanism.

Moreover, since $Q^2$ can be large, electroproduction is
a better process
from which to test the color-octet mechanism and to extract the NRQCD
long distance matrix elements than photoproduction.
The latter process lacks any large scale other than the charm
quark mass, and consequently, perturbative corrections to leading
order calculations are large.
In addition, nonperturbative effects, such as higher twist corrections to
the parton model, are less effectively
suppressed in photoproduction than in electroproduction at large $Q^2$.

The rest of the paper is organized as follows.
In Sec. II, we give the cross section formula of $\eta_c$ electroproduction in
the color-octet model. In Sec. III, we give the numerical results about the
production rates of $\eta_c$ at the energy range relevant to HERA collider.
The summary and discussions are given in Sec. IV.

\section{ $\eta_c$ ELECTROPRODUCTION IN THE COLOR OCTET MODEL}

The power of the NRQCD formalism stems from the fact that factorization
formulas for observables are expansions in the small parameter $v$ ,where
$v$ is the average relative velocity of the heavy quark and antiquark in
quarkonium bound state.
For charmonium $v^2\sim 0.3$ , and for bottomonium $v^2\sim 0.1$ .
NRQCD velocity-scaling rules\cite{glcuk} allow us to estimate the relative size of
various NRQCD matrix elements.
This information, along with the dependence of the short-distance coefficients
on $\alpha_s$ and $\alpha$, permits us to decide which terms must be retained
in expressions for observables to reach a given level of accuracy.
At low orders, factorization formulas involve only a few matrix elements, so
several observables can be related by a small of parameters.

In NRQCD the Fock state expansion for $\eta_c$ is
\begin{eqnarray}
\label{zankai}
\nonumber
|\eta_c \rangle
&=&O(1)|c\bar{c}({}^1S_{0},\b{1}) \rangle +
O(v)|c\bar{c}({}^1P_{1},\b{8})g\rangle +
    O(v^2)|c\bar{c}({}^3S_1,\b{8}~ or~ \b{1})gg\rangle\\
   && +O(v^2)|c\bar{c}({}^1S_0,\b{8}~ or~ \b{1})gg\rangle+\cdots.
\end{eqnarray}
For the production of $\eta_c$, the 
contributions to the NRQCD matrix elements 
from the last three terms of the above expansion are the same
order of $v^2$ according to the NRQCD velocity scaling rules. They are all
suppressed by $v^4$ compared to the contribution from the first term (the
color-singlet contribution).
However, the color-singlet contributions to $\eta_c$ production
vanish in the leading order and the
next to leading order $\gamma^* g$ fusion processes (see the following).
So, the production of $\eta_c$ is purely a color-octet process,
even to the next to leading order of QCD calculations.
Therefore, the $\eta_c$ production processes such
as at the HERA, will provide an important test for the color-octet production
mechanism.
Furthermore, we will show by the following calculations, the dominant contribution
to $\eta_c$ production comes from
the last term (color octet) of the Fock state expansion in Eq.
(\ref{zankai}). For this term, the associated color-octet production matrix
element $\langle {\cal O}_8^{\eta_c}({}^1S_0)\rangle $ can be related to the matrix element
$\langle {\cal O}_8^{\psi}({}^3S_1)\rangle $ by the heavy quark symmetry. And the latter
color-octet matrix element is important for the color-octet explanation of
the prompt $J/\psi$ surplus production at the Tevatron\cite{surplus,s1}.
So, the measurement of $\eta_c$ production processes
is closely associated with the 
test of the color-octet gluon fragmentation mechanism proposed in
\cite{surplus}.

The complete $ep\rightarrow \eta_c X$ cross secion can be written as
\begin{equation}
\label{ep}
\frac{d\sigma_{ep\rightarrow \eta_c X}(s)}{dydQ^2}=\Gamma (Q^2,y)d\sigma_{\gamma^*p\rightarrow \eta_c X}(W^2).
\end{equation}
Here $\Gamma (Q^2,y)$ is the
\begin{equation}
\label{gamma}
\Gamma (Q^2,y)=\frac{\alpha}{2\pi Q^2y}[1+(1-y)^2],
\end{equation}
and
\begin{equation}
\sigma(\gamma^*p\rightarrow \eta_c X)=\sigma_T(\gamma^*p\rightarrow \eta_c X)+
\sigma_L(\gamma^*p\rightarrow \eta_c X),
\end{equation}
where $\sigma_T$ and $\sigma_L$ are the cross sections of virtual photon
processes for the transverse
polarized photon and longitudinal polarized photon respectively.

According to the NRQCD  factorization formalism\cite{bbl}, the production process
$\gamma^*+ g\rightarrow \eta_c+X$ can be expressed as the following form,
\begin{equation} \label{xs}
d\sigma(\gamma^*+ g\rightarrow \eta_c+X)=\sum\limits_n F(\gamma^*+g\rightarrow n+X)\langle {\cal
O}_n^{\eta_c}\rangle .
\end{equation}
Here, $n$ denotes the $c\bar c$ pair configuration in the expansion terms of 
Eq. (\ref{zankai}) (including angular momentum $^{2S+1}L_J$ and color index
1 or 8). $F(\gamma^*+g\rightarrow n+X)$ is the short distance coefficient for the subprocess
$\gamma^* +g\rightarrow n+X$. $\langle {\cal O}_n^{\eta_c}\rangle $ is the long distance
non-perturbative matrix element which represents the probability of the
$c \bar c$ pair in $n$ configuration evolving into the physical state
$\eta_c$. The short distance coefficient $F$ can be calculated by using
perturbative QCD in expansion of powers of $\alpha_s$. The long distance
matrix elements are still not available from the first principles at
present. However, the relative importance of the contributions from
different terms in Eq. (\ref{xs}) can be estimated by using the NRQCD velocity
scaling rules.

From Eq. (\ref{zankai}), we can see that
the color-singlet contribution to the production of $\eta_c$
is at the leading order in $v^2$.
The associated short distance coefficient is given by the subprocess
\begin{equation}
\gamma^* g\rightarrow c\bar{c}({}^1S_{0},\b{1}) +g.
\end{equation}
This process occurs at the next to leading order in $\alpha_s$ for the $\gamma^* g$
fusion processes.
However, there is no contribution from this process,
because it violates $C$(charge) parity conservation.
(The $C$ parities
of the two gluon system (constrained in color-singlet) and the $c\bar c$ pair in
$({}^1S_{0},\b{1})$ state are both $+1$, while the $C$ parity of the photon
is $-1$).

The color-octet contributions to the $\eta_c$ production come from
the leading order and the next to leading order $\gamma^* g$ fusion processes
as shown in Fig.1.
At the leading order in $\alpha_s$, the subprocess is $2\rightarrow 1$ (Fig.1(a)),
\begin{equation}
\gamma^* g\rightarrow c\bar{c}({}^1S_0,\b{8}).
\end{equation}
For this process, we readily have
\begin{equation}
\sigma(\gamma^*+g\rightarrow c\bar c({}^1S_0,\b{8})\rightarrow\eta_c)=\frac{\pi^3e_c^2\alpha\alpha_s}
        {m_c^3}\delta(\hat{s}-4m_c^2-Q^2)\langle {\cal O}_8^{\eta_c}({}^1S_0)\rangle ,
\end{equation}
where $\hat{s}$ is the invariant mass squared of the partonic process. $m_c$ is
the charm quark mass, and we have approximated the charmonium bound state
mass of $\eta_c$ by $2m_c$.

At the next to leading order in $\alpha_s$, the subprocesses are $2\rightarrow 2$ (Fig.1(b)
and Fig.1(c)). These processes contain the contributions from
\begin{eqnarray}
\label{proc}
\gamma^* +g&\rightarrow& c\bar{c}({}^1S_0,\b{8})+g,\\
\label{proc1}
\gamma^* +g&\rightarrow& c\bar{c}({}^3S_1,\b{8})+g,\\
\label{proc2}
\gamma^* +g&\rightarrow& c\bar{c}({}^1P_1,\b{8})+g.
\end{eqnarray}
The cross sections for these $2\rightarrow 2$ subprocesses can be expressed
as the following form,
\begin{equation}
\label{fs}
\frac{d\sigma}{d\hat{t}}(\gamma^*+g\rightarrow \eta_c +X)=\frac{1}{16\pi\hat{s}^2}F({}^{2S+1}L_J^{(8)})
\langle {\cal O}_8^{\eta_c}({}^{2S+1}L_J)\rangle ,
\end{equation}
where $\hat{t}=(z-1)\hat{s}$, and $z$ is defined as $z=p\cdot k_{\eta_c}/
p\cdot k_\gamma^*$ with $p$, $k_{\eta_c}$, $k_{\gamma^*}$ being the momenta of the proton,
the outgoing $\eta_c$ and the incident photon respectively.

It is convenient to use the {\it helicity amplitude} method to calculate 
the cross section formulas for the virtual photon processes Eqs.(\ref{proc}-\ref{proc2}).
To the polarization vectors of the virtual photon, we choose
\begin{eqnarray}
\label{e1}
e_T = \frac{1}{\sqrt{2}}(0,1,\pm i,0), ~~e_L = \frac{1}{Q}p_1 + \frac{Q}{p_1\cdot p_2}p_2,
\end{eqnarray}
for the transverse and longitudinal polarized photons respectively.
For the incident and the outgoing gluons, 
following \cite{ham}, we choose their polarization vectors as
\begin{eqnarray}
\label{geprc}
\label{e2}
\not\! e_2^{(\pm)}=N_e[\not\! p_2\not\! p_3\not\! q(1\mp\gamma_5)+
        \not\! q\not\! p_3\not\! p_2(1\pm\gamma_5)],\\
\label{e3}
\not\! e_3^{(\pm)}=N_e[\not\! p_3\not\! q\not\! p_2(1\mp\gamma_5)+
        \not\! p_2\not\! q\not\! p_3(1\pm\gamma_5)].
\end{eqnarray}
Where $q = p_1 + \frac{Q^2}{2p_1\cdot p_2}p_2$ , $p_1^2=-Q^2$ and $p_1$ is the momenta for the incident photon.
$p_2,~p_3$ and $e_2,~e_3$ are the momenta and the 
polarization vectors for the incident gluon and outgoing 
gluon respectively.
The normalization factor $N_e$ is
\begin{equation}
N_e=\frac{1}{\sqrt{2(Q^2M^2 + \hat s \hat u)\hat t}},
\end{equation}
where $M$ is the mass of $\eta_c$, the Mandelstam invariants $\hat s,~\hat t,~\hat u$ are defined as
\begin{equation}
\hat s=(p_1+p_2)^2,~~~~\hat t=(p_2-p_3)^2,~~~~\hat u=(p_1-p_3)^2,
\end{equation}
and they satisfy the relation
$$\hat s+\hat t+\hat u=M^2 - Q^2 = 4m_c^2 - Q^2.$$

With these definitions for the polarization vectors of the photon and gluons
Eqs.~(\ref{e1}, \ref{e2}, \ref{e3}), 
the calculations of the helicity amplitudes are straightforward.
The expressions for $F$s in Eq.(\ref{fs}) are given in the Appendix.

\section{Numerical results}

For the numerical evaluation, we choose $m_c=1.5~GeV$, and set the
renormalization scale and the factorization scale both equal to
$\mu^2=(2m_c)^2+Q^2$.
For the parton distribution function of the proton, we use the
Gl\"uck-Reya-Vogt (GRV) leading order (LO) parametrization\cite{grv}.

Because there is no color-singlet contributions to $\eta_c$ electroproduction
up to the next to leading order perturbative calculations, we present in
the following the numerical results only coming from the color-octet
contributions.
For the color-octet $\eta_c$ production, by using the heavy quark spin
symmetry, we estimate the associated
color-octet matrix element $\langle {\cal O}_8^{\eta_c}({}^1S_0)\rangle $ to be
\begin{equation}
\langle {\cal O}_8^{\eta_c}({}^1S_0)\rangle \approx \langle {\cal O}_8^{\psi}({}^3S_1)\rangle =1.06\times 10^{-2} GeV^3.
\end{equation}
The value of the color-octet matrix element $\langle {\cal O}_8^{J/\psi
}({}^3S_1)\rangle$ follows the fitted value in\cite{benek} by comparing the
theoretical prediction of direct $J/\psi$ production to the experimental
data at the Tevatron. 
For the other two color-octet matrix elements, we estimate their values
by using the naive NRQCD velocity scaling rules,
\begin{eqnarray}
\langle {\cal O}_8^{\eta_c}({}^3S_1)\rangle &\approx &\langle {\cal O}_8^{\eta_c}({}^1S_0)\rangle ,\\
\langle {\cal O}_8^{\eta_c}({}^1P_1)\rangle &\approx &\langle {\cal O}_8^{\eta_c}({}^1S_0)\rangle .
\end{eqnarray}

In Fig.2, We first display the contribtuion from $2\rightarrow 1$ partonic subprocess
to the electroproduction of $\eta_c$, in which the intermediate color-octet
state is ${}^1S_0^{(8)}$.
The $2\rightarrow 1$ process contributes to the
$\eta_c$ production in the forward direction ($z\sim 1$).
In $e+p\rightarrow e+\eta_c+X$, we study the inelastic $\eta_c$ production 
in the kinematic range of $4<Q^2<80GeV^2$ and $40<W<180GeV^2$ for HERA
experiments.

Next we study the $2\rightarrow 2$ subprocesses contributions to the
electroproduction of $\eta_c$.
First, we consider the virtual photon-proton processes:
$\gamma^*p\rightarrow \eta_c X$.
In Fig.3, we plot the $z$ distributions of the cross sections
$\sigma_{tot}(\gamma^*p)=\sigma_T(\gamma^*p)+\sigma_L(\gamma^*p)$ of
$\eta_c$ production in the virtual photon-proton collisions at different
values of $Q^2$.
Relevant to the HERA energy range, we choose the photon-proton c.m. energy
$W=100GeV$ and select $Q^2=4,~10,~40GeV^2$ to see the change of the
cross sections with $Q^2$.
For comparison, we also show the photoproduction cross section in this figure
($Q^2=0$).
From these plots, we can see that the shapes of the curves for the individual
channel contributions do not change distinctly with $Q^2$, and their 
relative importances also do not change.
More important, ${}^1S_0^{(8)}$ channel always dominates over the other two
channels, which is the same as that in the photoproduction processes.

In Fig.4, we show the ratio $\sigma_L(\gamma^*p)/\sigma_T(\gamma^*p)$ as the function
of $z$ for $Q^2=10GeV^2$ and $40GeV^2$ respectively. We can see that this ratio increases
in all channels as $Q^2$ increases. In particular, for the ${}^3S_1^{(8)}$ channel, as $z\rightarrow 1$,
the ratio $\sigma_L(\gamma^*p)/\sigma_T(\gamma^*p)$ can reach a value over 4.4 for $Q^2=40GeV^2$.
This is a dictinct feature of the virtual photon with $Q^2>>m_c^2$.
But the total cross section $\sigma_{tot}(\gamma^*p)=\gamma_L(\gamma^*p)+\sigma_T(\gamma^*p)$
coming from the ${}^3S_1^{(8)}$ channel decreases. The total cross section is alwayse dominated
by the ${}^1S_0^{(8)}$ channel for all value of $Q^2$.

In Fig. 5, we show the cross sections of $e+p\rightarrow e+\eta_c+X$ at
HERA. We plot the differential cross sections $d\sigma/dz$ as a function of
$z$ for two cases for the integration region of $Q^2$: (a) $4GeV^2<Q^2<80GeV^2$;
(b) $20GeV^2<Q^2<80GeV^2$. 

From the above numerical results, we see that the contribution from the ${}^1S_0^{(8)}$
subprocess dominates the $\eta_c$ electroproduction.
The associated color-octet matrix
element $\langle {\cal O}_8^{\eta_c}({}^1S_0)\rangle $
can be related to the color-octet matrix element
$\langle {\cal O}_8^{J/\psi }({}^3S_1)\rangle$ by the heavy quark spin
symmetry.
The value of the later matrix element originates from
the fitting to the large transverse momentum $J/\psi$ production
at the Fermilab Tevatron\cite{benek}, which
can be viewed as a reliable estimate.
This is because the color-octet matrix element $\langle {\cal O}_8^{J/\psi }({}^3S_1)\rangle$
dominantly contributes in the large $p_T$ production
\cite{surplus,s1,explain},
where the initial and final state
gluon radiation effects are much smaller\cite{explain}.
So, the $\eta_c$ electroproduction together with its photoproduction
can provide an important test for the color-octet production mechanism.

We note that the next to leading order QCD perturbative
corrections are large for $J/\psi$ photoproduction at
HERA energy ranges\cite{MK,nlo,photo3}.
These corrections change the normalization of the production rate
by a factor of about $2$.
$\eta_c$ photoproduction and electroproduction may also
be affected by these corrections.
However, the main features in testing the color-singlet and
color-octet mechanisms mentioned above for $\eta_c$ production will not be changed.

For the experimental observation of $\eta_c$, we suggest using all main
decay channels of $\eta_c$ such as $\eta\pi\pi$, $\eta'\pi\pi$, $K\bar{K}\pi$,
$\pi\pi\pi\pi$, $K\bar{K}\pi\pi$, $K\bar{K}K\bar{K}$, as well as
$\gamma\gamma$ to detect $\eta_c$.
The increase of luminosity at HERA offers the possibility of
experimental investigation of eletrophotoproduction of $\eta_c$.
In this connection, we note that photoproduction of the $D_s$ meson
has been observed through the $D_s\rightarrow\phi\pi$ channel at HERA\cite{JB}.
According to our calculation, the photoproduction cross section of $\eta_c$ is
of the same order as the observed $D_s$.
The branching ratio of $\eta_c\rightarrow K\bar{K}\pi$ or $\eta_c\rightarrow\phi\phi$
is comparable to that of $D_s\rightarrow\phi\pi$.
Therefore, the experimental observation of photoproduction and leptoproduction
of $\eta_c$ might be possible through $\eta_c\rightarrow K\bar{K}\pi,\phi\phi$
decay modes in the near future.

\section{summary}
We have calculated in this paper the inelastic $\eta_c$ production
in the electroproduction processes under the NRQCD
factorization formalism.
Same as the photoproduction of $\eta_c$ \cite{photo-etac},
the color-singlet contributions to the electroproduction also vanish in
the leading order and the next to leading order in perturbative QCD.
The dominant contribution comes from the
color-octet ${}^1S_0^{(8)}$ subprocess, for which the associated color-octet
matrix element is related to the direct production
of $J/\psi$ at large transverse momentum at the Fermilab Tevatron.
The measurement of this process at HERA can be viewed as another independent
test for the color-octet production mechanism,
and complement to the study of its photoproduction.

\acknowledgments
This work was supported in part by the National Natural Science Foundation
of China, the State Education Commission of China, and the State
Commission of Science and Technology of China.

\newpage
\appendix
\section*{}
In this appendix, we list the cross sections for the different partonic processes.

\noindent
For ${}^1S_0$,${}^3S_1$ and ${}^1P_1$, the cross sections:
\begin{eqnarray}
\nonumber
F({}^3S_1^{(8)})&=&\frac{80(4\pi)^3e_c^2\alpha\alpha_s^2}{3( \hat{s}+\hat{t}+Q^2)
^2(\hat{s}+\hat{u}+2Q^2)^2(\hat{t}+\hat{u}+Q^2)^2(Q^2+s)^2}
[3Q^{14}+13Q^{12}\hat{s}+11Q^{12}\hat{t}\\
\nonumber
&&+7Q^{12}\hat{u} +22Q^{10}\hat{s}^2+40Q^{10}\hat{s}\hat{t}+30Q^{10}\hat{s}\hat{u}+16Q^{10}\hat{t}^2+18Q^{10}\hat{t}\hat{u}+5Q^{10}\hat{u}^2\\
\nonumber
&&+18Q^8\hat{s}^3 +55Q^8\hat{s}^2\hat{t}+50Q^8\hat{s}^2\hat{u}+51Q^8\hat{s}\hat{t}^2+64Q^8\hat{s}\hat{t}\hat{u}+21Q^8\hat{s}\hat{u}^2+12Q^8\hat{t}^3\\
\nonumber
&&+19Q^8\hat{t}^2\hat{u} +8Q^8\hat{t}\hat{u}^2+Q^8\hat{u}^3+7Q^6\hat{s}^4+35Q^6\hat{s}^3\hat{t}+40Q^6\hat{s}^3\hat{u}+61Q^6\hat{s}^2\hat{t}^2\\
\nonumber
&&+85Q^6\hat{s}^2\hat{t}\hat{u} +34Q^6\hat{s}^2\hat{u}^2+31Q^6\hat{s}\hat{t}^3+56Q^6\hat{s}\hat{t}^2\hat{u}+27Q^6\hat{s}\hat{t}\hat{u}^2+4Q^6\hat{s}\hat{u}^3+6Q^6\hat{t}^4\\
\nonumber
&&+11Q^6\hat{t}^3\hat{u} +7Q^6\hat{t}^2\hat{u}^2+Q^6\hat{t}\hat{u}^3 + Q^4\hat{s}^5+10Q^4\hat{s}^4\hat{t}+15Q^4\hat{s}^4\hat{u}+34Q^4\hat{s}^3\hat{t}^2\\
\nonumber
&&+51Q^4\hat{s}^3\hat{t}\hat{u} +26Q^4\hat{s}^3\hat{u}^2+27Q^4\hat{s}^2\hat{t}^3+58Q^4\hat{s}^2\hat{t}^2\hat{u}+33Q^4\hat{s}^2\hat{t}\hat{u}^2+6Q^4\hat{s}^2\hat{u}^3\\
\nonumber
&&+8Q^4\hat{s}\hat{t}^4 +21Q^4\hat{s}\hat{t}^3\hat{u}+17Q^4\hat{s}\hat{t}^2\hat{u}^2+3Q^4\hat{s}\hat{t}\hat{u}^3+2Q^4\hat{t}^5+4Q^4\hat{t}^4\hat{u}+3Q^4\hat{t}^3\hat{u}^2\\
\nonumber
&&+Q^4\hat{t}^2\hat{u}^3 +Q^2\hat{s}^5\hat{t}+2Q^2\hat{s}^5\hat{u}+9Q^2\hat{s}^4\hat{t}^2+13Q^2\hat{s}^4\hat{t}\hat{u}+9Q^2\hat{s}^4\hat{u}^2+9Q^2\hat{s}^3\hat{t}^3\\
\nonumber
&&+24Q^2\hat{s}^3\hat{t}^2\hat{u} +17Q^2\hat{s}^3\hat{t}\hat{u}^2+4Q^2\hat{s}^3\hat{u}^3+2Q^2\hat{s}^2\hat{t}^4 + 11Q^2\hat{s}^2\hat{t}^3\hat{u}+13Q^2\hat{s}^2\hat{t}^2\hat{u}^2\\
\nonumber
&&+3Q^2\hat{s}^2\hat{t}\hat{u}^3 +2Q^2\hat{s}\hat{t}^4\hat{u}+4Q^2\hat{s}\hat{t}^3\hat{u}^2+2Q^2\hat{s}\hat{t}^2\hat{u}^3+\hat{s}^5\hat{t}^2+\hat{s}^5\hat{t}\hat{u}+\hat{s}^5\hat{u}^2+\hat{s}^4\hat{t}^3\\
&&+3\hat{s}^4\hat{t}^2\hat{u} +3\hat{s}^4\hat{t}\hat{u}^2+\hat{s}^4\hat{u}^3+\hat{s}^3\hat{t}^3\hat{u}+3\hat{s}^3\hat{t}^2\hat{u}^2+\hat{s}^3\hat{t}\hat{u}^3+\hat{s}^2\hat{t}^3\hat{u}^2+\hat{s}^2\hat{t}^2\hat{u}^3].
\end{eqnarray}

\noindent
\begin{eqnarray}
\nonumber
F({}^1S_0^{(8)})&=&\frac{24(4\pi)^3e_c^2\alpha\alpha_s^2}{(Q^2+\hat{s}+\hat{t})^2(2Q^2+\hat{s}+\hat{u})^2(Q^2+\hat{s})^2(Q^2+\hat{t}+\hat{u}
) ^2\hat{t}}
[9Q^{16} + 45Q^{14}\hat{s} + 49Q^{14}\hat{t}\\
\nonumber
&& + 27Q^{14}\hat{u} + 96Q^{12}\hat{s}^2 + 174Q^{12}\hat{s}\hat{t} + 123Q^{12}\hat{s}\hat{u} + 94Q^{12}\hat{t}^2 + 136Q^{12}\hat{t}\hat{u} + 33Q^{12}\hat{u}^2\\
\nonumber
&& + 114Q^{10}\hat{s}^3 + 256Q^{10}\hat{s}^2\hat{t} + 234Q^{10}\hat{s}^2\hat{u} + 250Q^{10}\hat{s}\hat{t}^2 + 430Q^{10}\hat{s}\hat{t}\hat{u} + 135Q^{10}\hat{s}\hat{u}^2\\
\nonumber
&& + 80Q^{10}\hat{t}^3 + 202Q^{10}\hat{t}^2\hat{u} + 145Q^{10}\hat{t}\hat{u}^2 + 21Q^{10}\hat{u}^3 + 82Q^8\hat{s}^4+ 206Q^8\hat{s}^3\hat{t}\\
\nonumber
&& + 242Q^8\hat{s}^3\hat{u} + 273Q^8\hat{s}^2\hat{t}^2 + 538Q^8\hat{s}^2\hat{t}\hat{u} + 222Q^8\hat{s}^2\hat{u}^2 + 172Q^8\hat{s}\hat{t}^3 + 462Q^8\hat{s}\hat{t}^2\hat{u}\\
\nonumber
&& + 406Q^8\hat{s}\hat{t}\hat{u}^2 + 77Q^8\hat{s}\hat{u}^3 + 31Q^8\hat{t}^4 + 112Q^8\hat{t}^3\hat{u} + 145Q^8\hat{t}^2\hat{u}^2 + 70Q^8\hat{t}\hat{u}^3 + 7Q^8\hat{u}^4\\
\nonumber
&& + 36Q^6\hat{s}^5 + 98Q^6\hat{s}^4\hat{t} + 148Q^6\hat{s}^4\hat{u} + 157Q^6\hat{s}^3\hat{t}^2 + 352Q^6\hat{s}^3\hat{t}\hat{u} + 188Q^6\hat{s}^3\hat{u}^2\\
\nonumber
&& + 153Q^6\hat{s}^2\hat{t}^3 + 413Q^6\hat{s}^2\hat{t}^2\hat{u} + 420Q^6\hat{s}^2\hat{t}\hat{u}^2 + 108Q^6\hat{s}^2\hat{u}^3 + 63Q^6\hat{s}\hat{t}^4 + 194Q^6\hat{s}\hat{t}^3\hat{u}\\
\nonumber
&& + 275Q^6\hat{s}\hat{t}^2\hat{u}^2 + 172Q^6\hat{s}\hat{t}\hat{u}^3 + 23Q^6\hat{s}\hat{u}^4 + 5Q^6\hat{t}^5 + 21Q^6\hat{t}^4\hat{u} + 39Q^6\hat{t}^3\hat{u}^2\\
\nonumber
&& + 35Q^6\hat{t}^2\hat{u}^3 + 13Q^6\hat{t}\hat{u}^4 + Q^6\hat{u}^5 + 9Q^4\hat{s}^6 + 26Q^4\hat{s}^5\hat{t} + 54Q^4\hat{s}^5\hat{u} + 45Q^4\hat{s}^4\hat{t}^2\\
\nonumber
&& + 132Q^4\hat{s}^4\hat{t}\hat{u} + 87Q^4\hat{s}^4\hat{u}^2 + 60Q^4\hat{s}^3\hat{t}^3 + 191Q^4\hat{s}^3\hat{t}^2\hat{u} + 208Q^4\hat{s}^3\hat{t}\hat{u}^2 + 72Q^4\hat{s}^3\hat{u}^3\\
\nonumber
&& + 43Q^4\hat{s}^2\hat{t}^4 + 134Q^4\hat{s}^2\hat{t}^3\hat{u} + 181Q^4\hat{s}^2\hat{t}^2\hat{u}^2 + 140Q^4\hat{s}^2\hat{t}\hat{u}^3 + 27Q^4\hat{s}^2\hat{u}^4 + 10Q^4\hat{s}\hat{t}^5\\
\nonumber
&& + 33Q^4\hat{s}\hat{t}^4\hat{u} + 50Q^4\hat{s}\hat{t}^3\hat{u}^2 + 53Q^4\hat{s}\hat{t}^2\hat{u}^3 + 28Q^4\hat{s}\hat{t}\hat{u}^4 + 3Q^4\hat{s}\hat{u}^5 + Q^2\hat{s}^7 + 3Q^2\hat{s}^6\hat{t}\\
\nonumber
&& + 11Q^2\hat{s}^6\hat{u} + 5Q^2\hat{s}^5\hat{t}^2 + 26Q^2\hat{s}^5\hat{t}\hat{u} + 21Q^2\hat{s}^5\hat{u}^2 + 9Q^2\hat{s}^4\hat{t}^3 + 41Q^2\hat{s}^4\hat{t}^2\hat{u}\\
\nonumber
&& + 55Q^2\hat{s}^4\hat{t}\hat{u}^2 + 23Q^2\hat{s}^4\hat{u}^3 + 11Q^2\hat{s}^3\hat{t}^4 + 34Q^2\hat{s}^3\hat{t}^3\hat{u} + 57Q^2\hat{s}^3\hat{t}^2\hat{u}^2 + 44Q^2\hat{s}^3\hat{t}\hat{u}^3 \\
\nonumber
&&+ 13Q^2\hat{s}^3\hat{u}^4 + 5Q^2\hat{s}^2\hat{t}^5 + 13Q^2\hat{s}^2\hat{t}^4\hat{u} + 23Q^2\hat{s}^2\hat{t}^3\hat{u}^2 + 21Q^2\hat{s}^2\hat{t}^2\hat{u}^3 + 17Q^2\hat{s}^2\hat{t}\hat{u}^4 \\
\nonumber
&&+ 3Q^2\hat{s}^2\hat{u}^5 + \hat{s}^7\hat{u} + 2\hat{s}^6\hat{t}\hat{u} + 2\hat{s}^6\hat{u}^2 + 3\hat{s}^5\hat{t}^2\hat{u} + 6\hat{s}^5\hat{t}\hat{u}^2 + 3\hat{s}^5\hat{u}^3 + 2\hat{s}^4\hat{t}^3\hat{u} + 6\hat{s}^4\hat{t}^2\hat{u}^2\\
&& + 6\hat{s}^4\hat{t}\hat{u}^3 + 2\hat{s}^4\hat{u}^4 + \hat{s}^3\hat{t}^4\hat{u} + 2\hat{s}^3\hat{t}^3\hat{u}^2 + 3\hat{s}^3\hat{t}^2\hat{u}^3 + 2\hat{s}^3\hat{t}\hat{u}^4 + \hat{s}^3\hat{u}^5].
\end{eqnarray}

\noindent
\begin{eqnarray}
\nonumber
F({}^1P_1^{(8)})&=&\frac{320(4\pi)^3e_c^2\alpha\alpha_s^2}{3(2m_c)^2(Q^2 + \hat{s}+\hat{t})
^3(2Q^2 + \hat{s}+\hat{u}) ^4(Q^2 + \hat{t}+\hat{u}
) ^3(Q^2+\hat{s})^2}\\
\nonumber
&&
[48Q^{22} + 336Q^{20}\hat{s} + 336Q^{20}\hat{t} + 240Q^{20}\hat{u} + 1048Q^{18}\hat{s}^2 + 1900Q^{18}\hat{s}\hat{t} + 1552Q^{18}\hat{s}\hat{u}\\
\nonumber
&&+ 922Q^{18}\hat{t}^2 + 1548Q^{18}\hat{t}\hat{u} + 520Q^{18}\hat{u}^2 + 1920Q^{16}\hat{s}^3 + 4716Q^{16}\hat{s}^2\hat{t} + 4432Q^{16}\hat{s}^2\hat{u} \\
\nonumber
&&+ 4304Q^{16}\hat{s}\hat{t}^2 + 7960Q^{16}\hat{s}\hat{t}\hat{u} + 3088Q^{16}\hat{s}\hat{u}^2 + 1410Q^{16}\hat{t}^3 + 3700Q^{16}\hat{t}^2\hat{u} \\
\nonumber
&& + 3044Q^{16}\hat{t}\hat{u}^2+ 640Q^{16}\hat{u}^3 + 2291Q^{14}\hat{s}^4 + 6795Q^{14}\hat{s}^3\hat{t} + 7356Q^{14}\hat{s}^3\hat{u}\\
\nonumber
&& + 8702Q^{14}\hat{s}^2\hat{t}^2 + 17713Q^{14}\hat{s}^2\hat{t}\hat{u} + 7994Q^{14}\hat{s}^2\hat{u}^2 + 5498Q^{14}\hat{s}\hat{t}^3 + 15470Q^{14}\hat{s}\hat{t}^2\hat{u}\\
\nonumber
&& + 14125Q^{14}\hat{s}\hat{t}\hat{u}^2 + 3468Q^{14}\hat{s}\hat{u}^3 + 1518Q^{14}\hat{t}^4 + 4898Q^{14}\hat{t}^3\hat{u} + 6220Q^{14}\hat{t}^2\hat{u}^2\\
\nonumber
&& + 3319Q^{14}\hat{t}\hat{u}^3 + 491Q^{14}\hat{u}^4 + 1863Q^{12}\hat{s}^5+ 6339Q^{12}\hat{s}^4\hat{t} + 7853Q^{12}\hat{s}^4\hat{u} \\
\nonumber
&&+ 10120Q^{12}\hat{s}^3\hat{t}^2 + 22468Q^{12}\hat{s}^3\hat{t}\hat{u} + 11838Q^{12}\hat{s}^3\hat{u}^2 + 9156Q^{12}\hat{s}^2\hat{t}^3 + 27460Q^{12}\hat{s}^2\hat{t}^2\hat{u}\\
\nonumber
&& + 27824Q^{12}\hat{s}^2\hat{t}\hat{u}^2 + 8050Q^{12}\hat{s}^2\hat{u}^3 + 4972Q^{12}\hat{s}\hat{t}^4 + 16826Q^{12}\hat{s}\hat{t}^3\hat{u} + 22970Q^{12}\hat{s}\hat{t}^2\hat{u}^2\\
\nonumber
&& + 13768Q^{12}\hat{s}\hat{t}\hat{u}^3 + 2411Q^{12}\hat{s}\hat{u}^4 + 1334Q^{12}\hat{t}^5 + 4676Q^{12}\hat{t}^4\hat{u} + 7100Q^{12}\hat{t}^3\hat{u}^2\\
\nonumber
&& + 5690Q^{12}\hat{t}^2\hat{u}^3 + 2177Q^{12}\hat{t}\hat{u}^4 + 241Q^{12}\hat{u}^5 + 1047Q^{10}\hat{s}^6 + 4042Q^{10}\hat{s}^5\hat{t}\\
\nonumber
&&+ 5628Q^{10}\hat{s}^5\hat{u} + 7609Q^{10}\hat{s}^4\hat{t}^2 + 18020Q^{10}\hat{s}^4\hat{t}\hat{u} + 11052Q^{10}\hat{s}^4\hat{u}^2\\
\nonumber
&& + 8639Q^{10}\hat{s}^3\hat{t}^3 + 27282Q^{10}\hat{s}^3\hat{t}^2\hat{u} + 30485Q^{10}\hat{s}^3\hat{t}\hat{u}^2 + 10452Q^{10}\hat{s}^3\hat{u}^3 + 6717Q^{10}\hat{s}^2\hat{t}^4\\
\nonumber
&& + 23967Q^{10}\hat{s}^2\hat{t}^3\hat{u} + 35078Q^{10}\hat{s}^2\hat{t}^2\hat{u}^2 + 23615Q^{10}\hat{s}^2\hat{t}\hat{u}^3 + 4955Q^{10}\hat{s}^2\hat{u}^4\\
\nonumber
&& + 3568Q^{10}\hat{s}\hat{t}^5 + 13176Q^{10}\hat{s}\hat{t}^4\hat{u} + 20989Q^{10}\hat{s}\hat{t}^3\hat{u}^2 + 18212Q^{10}\hat{s}\hat{t}^2\hat{u}^3 + 7977Q^{10}\hat{s}\hat{t}\hat{u}^4\\
\nonumber
&& + 1064Q^{10}\hat{s}\hat{u}^5 + 902Q^{10}\hat{t}^6 + 3520Q^{10}\hat{t}^5\hat{u} + 5917Q^{10}\hat{t}^4\hat{u}^2 + 5581Q^{10}\hat{t}^3\hat{u}^3\\
\nonumber
&& + 3079Q^{10}\hat{t}^2\hat{u}^4 + 869Q^{10}\hat{t}\hat{u}^5 + 74Q^{10}\hat{u}^6 + 402Q^8\hat{s}^7 + 1807Q^8\hat{s}^6\hat{t}\\
\nonumber
&& + 2736Q^8\hat{s}^6\hat{u} + 3946Q^8\hat{s}^5\hat{t}^2 + 9568Q^8\hat{s}^5\hat{t}\hat{u} + 6747Q^8\hat{s}^5\hat{u}^2 + 5199Q^8\hat{s}^4\hat{t}^3\\
\nonumber
&&+ 16875Q^8\hat{s}^4\hat{t}^2\hat{u} + 20452Q^8\hat{s}^4\hat{t}\hat{u}^2 + 8305Q^8\hat{s}^4\hat{u}^3 + 4895Q^8\hat{s}^3\hat{t}^4 + 18541Q^8\hat{s}^3\hat{t}^3\hat{u}\\
\nonumber
&& + 28923Q^8\hat{s}^3\hat{t}^2\hat{u}^2 + 21762Q^8\hat{s}^3\hat{t}\hat{u}^3 + 5520Q^8\hat{s}^3\hat{u}^4 + 3687Q^8\hat{s}^2\hat{t}^5 + 14625Q^8\hat{s}^2\hat{t}^4\hat{u}\\
\nonumber
&& + 24778Q^8\hat{s}^2\hat{t}^3\hat{u}^2 + 23241Q^8\hat{s}^2\hat{t}^2\hat{u}^3 + 11647Q^8\hat{s}^2\hat{t}\hat{u}^4 + 1906Q^8\hat{s}^2\hat{u}^5 + 1848Q^8\hat{s}\hat{t}^6\\
\nonumber
&& + 7722Q^8\hat{s}\hat{t}^5\hat{u} + 13765Q^8\hat{s}\hat{t}^4\hat{u}^2 + 13745Q^8\hat{s}\hat{t}^3\hat{u}^3 + 8329Q^8\hat{s}\hat{t}^2\hat{u}^4 + 2770Q^8\hat{s}\hat{t}\hat{u}^5\\
\nonumber
&& + 291Q^8\hat{s}\hat{u}^6 + 398Q^8\hat{t}^7 + 1804Q^8\hat{t}^6\hat{u} + 3523Q^8\hat{t}^5\hat{u}^2 + 3855Q^8\hat{t}^4\hat{u}^3 + 2547Q^8\hat{t}^3\hat{u}^4\\
\nonumber
&& + 998Q^8\hat{t}^2\hat{u}^5 + 202Q^8\hat{t}\hat{u}^6 + 13Q^8\hat{u}^7 + 101Q^6\hat{s}^8 + 567Q^6\hat{s}^7\hat{t} + 887Q^6\hat{s}^7\hat{u}\\
\nonumber
&& + 1458Q^6\hat{s}^6\hat{t}^2 + 3403Q^6\hat{s}^6\hat{t}\hat{u} + 2693Q^6\hat{s}^6\hat{u}^2 + 2141Q^6\hat{s}^5\hat{t}^3 + 6828Q^6\hat{s}^5\hat{t}^2\hat{u}\\
\nonumber
&& + 8693Q^6\hat{s}^5\hat{t}\hat{u}^2 + 4139Q^6\hat{s}^5\hat{u}^3 + 2154Q^6\hat{s}^4\hat{t}^4 + 8625Q^6\hat{s}^4\hat{t}^3\hat{u} + 14103Q^6\hat{s}^4\hat{t}^2\hat{u}^2\\
\nonumber
&& + 11713Q^6\hat{s}^4\hat{t}\hat{u}^3 + 3600Q^6\hat{s}^4\hat{u}^4 + 1868Q^6\hat{s}^3\hat{t}^5 + 8205Q^6\hat{s}^3\hat{t}^4\hat{u} + 14993Q^6\hat{s}^3\hat{t}^3\hat{u}^2\\
\nonumber
&& + 15220Q^6\hat{s}^3\hat{t}^2\hat{u}^3 + 8678Q^6\hat{s}^3\hat{t}\hat{u}^4 + 1769Q^6\hat{s}^3\hat{u}^5 + 1337Q^6\hat{s}^2\hat{t}^6 + 6118Q^6\hat{s}^2\hat{t}^5\hat{u}\\
\nonumber
&& + 11895Q^6\hat{s}^2\hat{t}^4\hat{u}^2 + 12811Q^6\hat{s}^2\hat{t}^3\hat{u}^3 + 8502Q^6\hat{s}^2\hat{t}^2\hat{u}^4 + 3332Q^6\hat{s}^2\hat{t}\hat{u}^5 + 445Q^6\hat{s}^2\hat{u}^6\\
\nonumber
&& + 570Q^6\hat{s}\hat{t}^7 + 2792Q^6\hat{s}\hat{t}^6\hat{u} + 5910Q^6\hat{s}\hat{t}^5\hat{u}^2 + 6989Q^6\hat{s}\hat{t}^4\hat{u}^3 + 5004Q^6\hat{s}\hat{t}^3\hat{u}^4\\
\nonumber
&& + 2210Q^6\hat{s}\hat{t}^2\hat{u}^5 + 550Q^6\hat{s}\hat{t}\hat{u}^6 + 45Q^6\hat{s}\hat{u}^7 + 98Q^6\hat{t}^8 + 514Q^6\hat{t}^7\hat{u} + 1187Q^6\hat{t}^6\hat{u}^2\\
\nonumber
&& + 1572Q^6\hat{t}^5\hat{u}^3 + 1287Q^6\hat{t}^4\hat{u}^4 + 646Q^6\hat{t}^3\hat{u}^5 + 183Q^6\hat{t}^2\hat{u}^6 + 24Q^6\hat{t}\hat{u}^7 + Q^6\hat{u}^8\\
\nonumber
&& + 15Q^4\hat{s}^9 + 121Q^4\hat{s}^8\hat{t} + 182Q^4\hat{s}^8\hat{u} + 379Q^4\hat{s}^7\hat{t}^2 + 791Q^4\hat{s}^7\hat{t}\hat{u} + 677Q^4\hat{s}^7\hat{u}^2\\
\nonumber
&& + 620Q^4\hat{s}^6\hat{t}^3 + 1823Q^4\hat{s}^6\hat{t}^2\hat{u} + 2317Q^4\hat{s}^6\hat{t}\hat{u}^2 + 1265Q^4\hat{s}^6\hat{u}^3 + 623Q^4\hat{s}^5\hat{t}^4\\
\nonumber
&& + 2536Q^4\hat{s}^5\hat{t}^3\hat{u} + 4166Q^4\hat{s}^5\hat{t}^2\hat{u}^2 + 3725Q^4\hat{s}^5\hat{t}\hat{u}^3 + 1376Q^4\hat{s}^5\hat{u}^4\\
\nonumber
&& + 498Q^4\hat{s}^4\hat{t}^5 + 2545Q^4\hat{s}^4\hat{t}^4\hat{u} + 4986Q^4\hat{s}^4\hat{t}^3\hat{u}^2 + 5449Q^4\hat{s}^4\hat{t}^2\hat{u}^3\\
\nonumber
&& + 3502Q^4\hat{s}^4\hat{t}\hat{u}^4 + 896Q^4\hat{s}^4\hat{u}^5 + 393Q^4\hat{s}^3\hat{t}^6 + 2150Q^4\hat{s}^3\hat{t}^5\hat{u} + 4713Q^4\hat{s}^3\hat{t}^4\hat{u}^2 \\
\nonumber
&&+ 5603Q^4\hat{s}^3\hat{t}^3\hat{u}^3 + 4117Q^4\hat{s}^3\hat{t}^2\hat{u}^4 + 1881Q^4\hat{s}^3\hat{t}\hat{u}^5 + 329Q^4\hat{s}^3\hat{u}^6 + 237Q^4\hat{s}^2\hat{t}^7\\
\nonumber
&& + 1343Q^4\hat{s}^2\hat{t}^6\hat{u} + 3229Q^4\hat{s}^2\hat{t}^5\hat{u}^2 + 4283Q^4\hat{s}^2\hat{t}^4\hat{u}^3 + 3395Q^4\hat{s}^2\hat{t}^3\hat{u}^4\\
\nonumber
&& + 1683Q^4\hat{s}^2\hat{t}^2\hat{u}^5 + 519Q^4\hat{s}^2\hat{t}\hat{u}^6 + 57Q^4\hat{s}^2\hat{u}^7 + 76Q^4\hat{s}\hat{t}^8 + 460Q^4\hat{s}\hat{t}^7\hat{u}\\
\nonumber
&& + 1205Q^4\hat{s}\hat{t}^6\hat{u}^2 + 1804Q^4\hat{s}\hat{t}^5\hat{u}^3 + 1670Q^4\hat{s}\hat{t}^4\hat{u}^4 + 955Q^4\hat{s}\hat{t}^3\hat{u}^5\\
\nonumber
&& + 320Q^4\hat{s}\hat{t}^2\hat{u}^6 + 55Q^4\hat{s}\hat{t}\hat{u}^7 + 3Q^4\hat{s}\hat{u}^8 + 10Q^4\hat{t}^9 + 60Q^4\hat{t}^8\hat{u} + 161Q^4\hat{t}^7\hat{u}^2\\
\nonumber
&& + 255Q^4\hat{t}^6\hat{u}^3 + 261Q^4\hat{t}^5\hat{u}^4 + 174Q^4\hat{t}^4\hat{u}^5 + 71Q^4\hat{t}^3\hat{u}^6 + 15Q^4\hat{t}^2\hat{u}^7 + Q^4\hat{t}\hat{u}^8\\
\nonumber
&& + Q^2\hat{s}^10 + 16Q^2\hat{s}^9\hat{t} + 21Q^2\hat{s}^9\hat{u} + 63Q^2\hat{s}^8\hat{t}^2 + 110Q^2\hat{s}^8\hat{t}\hat{u} + 97Q^2\hat{s}^8\hat{u}^2 \\
\nonumber
&&+ 118Q^2\hat{s}^7\hat{t}^3 + 302Q^2\hat{s}^7\hat{t}^2\hat{u} + 359Q^2\hat{s}^7\hat{t}\hat{u}^2 + 217Q^2\hat{s}^7\hat{u}^3 + 125Q^2\hat{s}^6\hat{t}^4\\
\nonumber
&& + 466Q^2\hat{s}^6\hat{t}^3\hat{u} + 715Q^2\hat{s}^6\hat{t}^2\hat{u}^2 + 659Q^2\hat{s}^6\hat{t}\hat{u}^3 + 286Q^2\hat{s}^6\hat{u}^4 + 84Q^2\hat{s}^5\hat{t}^5\\
\nonumber
&& + 461Q^2\hat{s}^5\hat{t}^4\hat{u} + 910Q^2\hat{s}^5\hat{t}^3\hat{u}^2 + 1032Q^2\hat{s}^5\hat{t}^2\hat{u}^3 + 735Q^2\hat{s}^5\hat{t}\hat{u}^4 + 235Q^2\hat{s}^5\hat{u}^5 \\
\nonumber
&&+ 47Q^2\hat{s}^4\hat{t}^6 + 344Q^2\hat{s}^4\hat{t}^5\hat{u} + 876Q^2\hat{s}^4\hat{t}^4\hat{u}^2 + 1148Q^2\hat{s}^4\hat{t}^3\hat{u}^3 + 949Q^2\hat{s}^4\hat{t}^2\hat{u}^4\\
\nonumber
&& + 501Q^2\hat{s}^4\hat{t}\hat{u}^5 + 117Q^2\hat{s}^4\hat{u}^6 + 26Q^2\hat{s}^3\hat{t}^7 + 206Q^2\hat{s}^3\hat{t}^6\hat{u} + 640Q^2\hat{s}^3\hat{t}^5\hat{u}^2\\
\nonumber
&& + 999Q^2\hat{s}^3\hat{t}^4\hat{u}^3 + 916Q^2\hat{s}^3\hat{t}^3\hat{u}^4 + 528Q^2\hat{s}^3\hat{t}^2\hat{u}^5 + 196Q^2\hat{s}^3\hat{t}\hat{u}^6 + 31Q^2\hat{s}^3\hat{u}^7 \\
\nonumber
&&+ 8Q^2\hat{s}^2\hat{t}^8 + 78Q^2\hat{s}^2\hat{t}^7\hat{u} + 289Q^2\hat{s}^2\hat{t}^6\hat{u}^2 + 542Q^2\hat{s}^2\hat{t}^5\hat{u}^3 + 599Q^2\hat{s}^2\hat{t}^4\hat{u}^4\\
\nonumber
&& + 402Q^2\hat{s}^2\hat{t}^3\hat{u}^5 + 161Q^2\hat{s}^2\hat{t}^2\hat{u}^6 + 38Q^2\hat{s}^2\hat{t}\hat{u}^7 + 3Q^2\hat{s}^2\hat{u}^8 + 10Q^2\hat{s}\hat{t}^8\hat{u} + 52Q^2\hat{s}\hat{t}^7\hat{u}^2\\
\nonumber
&& + 120Q^2\hat{s}\hat{t}^6\hat{u}^3 + 162Q^2\hat{s}\hat{t}^5\hat{u}^4 + 138Q^2\hat{s}\hat{t}^4\hat{u}^5 + 72Q^2\hat{s}\hat{t}^3\hat{u}^6 + 20Q^2\hat{s}\hat{t}^2\hat{u}^7 + 2Q^2\hat{s}\hat{t}\hat{u}^8\\
\nonumber
&& + \hat{s}^10\hat{t} + \hat{s}^10\hat{u} + 5\hat{s}^9\hat{t}^2 + 7\hat{s}^9\hat{t}\hat{u} + 6\hat{s}^9\hat{u}^2 + 11\hat{s}^8\hat{t}^3 + 24\hat{s}^8\hat{t}^2\hat{u} + 25\hat{s}^8\hat{t}\hat{u}^2\\
\nonumber
&& + 16\hat{s}^8\hat{u}^3 + 14\hat{s}^7\hat{t}^4 + 43\hat{s}^7\hat{t}^3\hat{u} + 57\hat{s}^7\hat{t}^2\hat{u}^2 + 51\hat{s}^7\hat{t}\hat{u}^3 + 25\hat{s}^7\hat{u}^4 + 11\hat{s}^6\hat{t}^5 + 46\hat{s}^6\hat{t}^4\hat{u}\\
\nonumber
&& + 78\hat{s}^6\hat{t}^3\hat{u}^2 + 84\hat{s}^6\hat{t}^2\hat{u}^3 + 64\hat{s}^6\hat{t}\hat{u}^4 + 25\hat{s}^6\hat{u}^5 + 5\hat{s}^5\hat{t}^6 + 30\hat{s}^5\hat{t}^5\hat{u} + 70\hat{s}^5\hat{t}^4\hat{u}^2 \\
\nonumber
&&+ 92\hat{s}^5\hat{t}^3\hat{u}^3 + 84\hat{s}^5\hat{t}^2\hat{u}^4 + 51\hat{s}^5\hat{t}\hat{u}^5 + 16\hat{s}^5\hat{u}^6 + \hat{s}^4\hat{t}^7 + 11\hat{s}^4\hat{t}^6\hat{u} + 38\hat{s}^4\hat{t}^5\hat{u}^2 + 70\hat{s}^4\hat{t}^4\hat{u}^3\\
\nonumber
&& +78\hat{s}^4\hat{t}^3\hat{u}^4 + 57\hat{s}^4\hat{t}^2\hat{u}^5 + 25\hat{s}^4\hat{t}\hat{u}^6 + 6\hat{s}^4\hat{u}^7 + 2\hat{s}^3\hat{t}^7\hat{u} + 11\hat{s}^3\hat{t}^6\hat{u}^2 + 30\hat{s}^3\hat{t}^5\hat{u}^3\\
\nonumber
&& + 46\hat{s}^3\hat{t}^4\hat{u}^4 + 43\hat{s}^3\hat{t}^3\hat{u}^5 + 24\hat{s}^3\hat{t}^2\hat{u}^6 + 7\hat{s}^3\hat{t}\hat{u}^7 + \hat{s}^3\hat{u}^8 + \hat{s}^2\hat{t}^7\hat{u}^2 + 5\hat{s}^2\hat{t}^6\hat{u}^3 + 11\hat{s}^2\hat{t}^5\hat{u}^4\\
&& + 14\hat{s}^2\hat{t}^4\hat{u}^5 + 11\hat{s}^2\hat{t}^3\hat{u}^6 + 5\hat{s}^2\hat{t}^2\hat{u}^7 + \hat{s}^2\hat{t}\hat{u}^8].
\end{eqnarray}

\newpage
\vskip 10mm
\centerline{\bf \large Figure Captions}
\vskip 1cm

FIG. 1. The Feynman diagrams for the production of $\eta_c$
at the leading order $(a)$ and the next to leading order $(b) and (c)$
$\gamma^* g$ fusion processes. All of these diagrams are for the
color-octet processes.

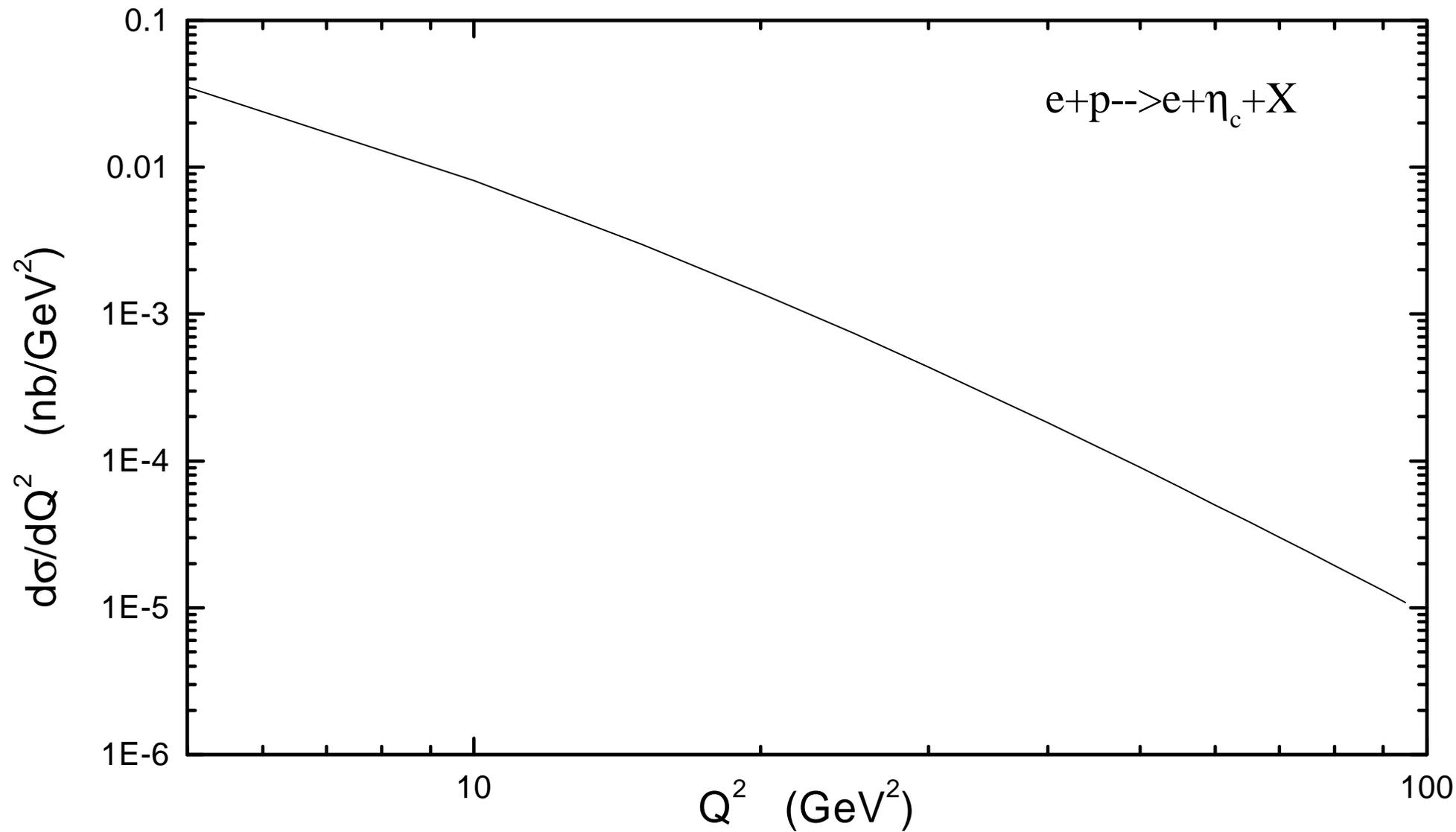
FIG. 2. The cross section $d\sigma /dQ^2$ for the process $e+p\rightarrow e+\eta_c+X$ in the forward direction ($z\sim 1$) as a function of $Q^2$. where $\sqrt{s_{ep}}=300GeV$.

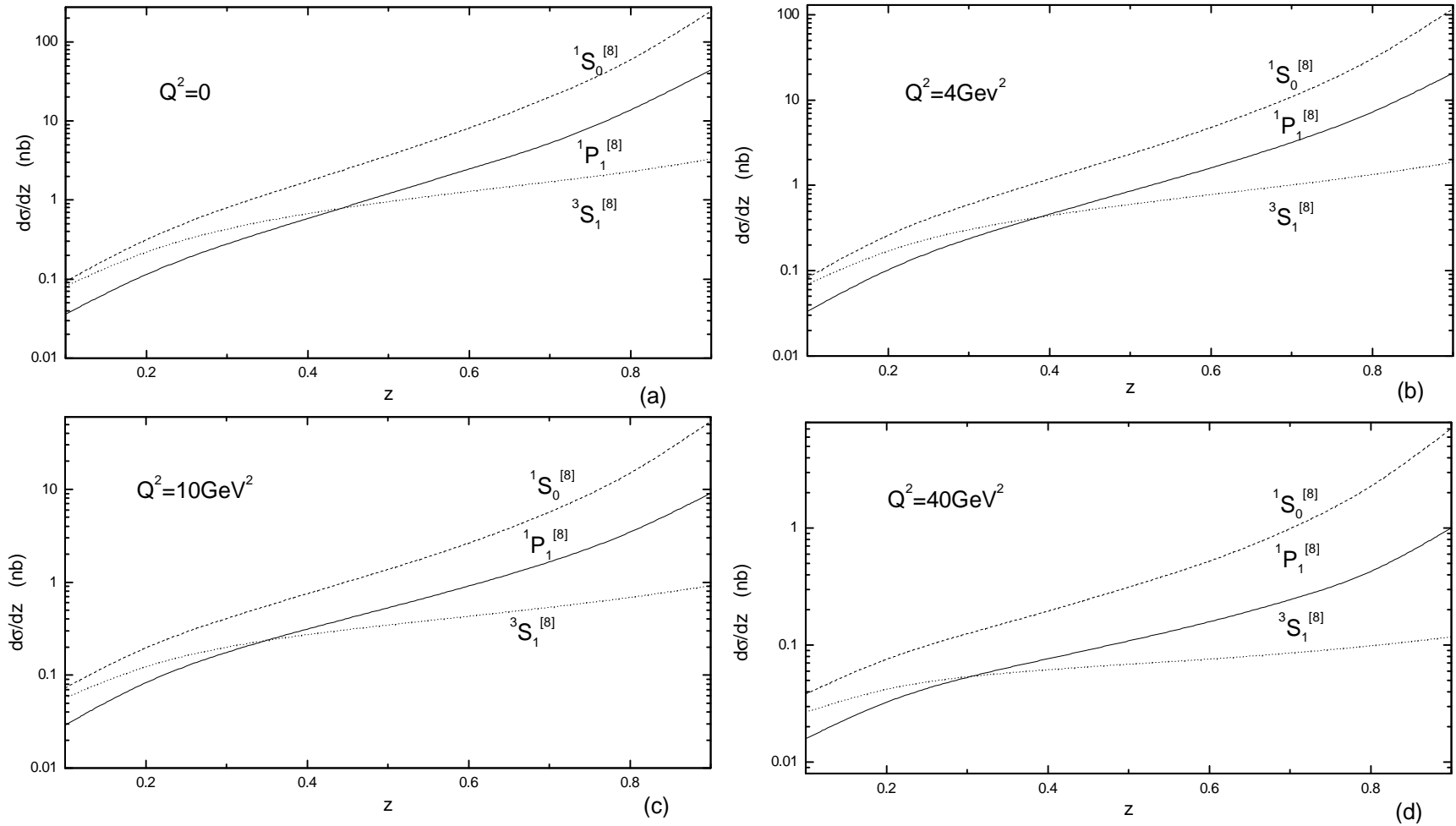
FIG. 3. The differential cross sections $d\sigma /dz$ for the process $\gamma^*
+p\rightarrow \eta_c+X$ at the HERA as a function of $z\equiv E_{\eta
_c}/E_\gamma^* $, where $W_{\gamma^* p}=100GeV$; (a) $Q^2=0GeV^2$;
(b) $Q^2=4GeV^2$; (c) $Q^2=10GeV^2$; (d) $Q^2=40GeV^2$.
The curves denote contributions from the color-octet $^1S_0$, $^1P_1$,
and $^3S_1$ respectively.

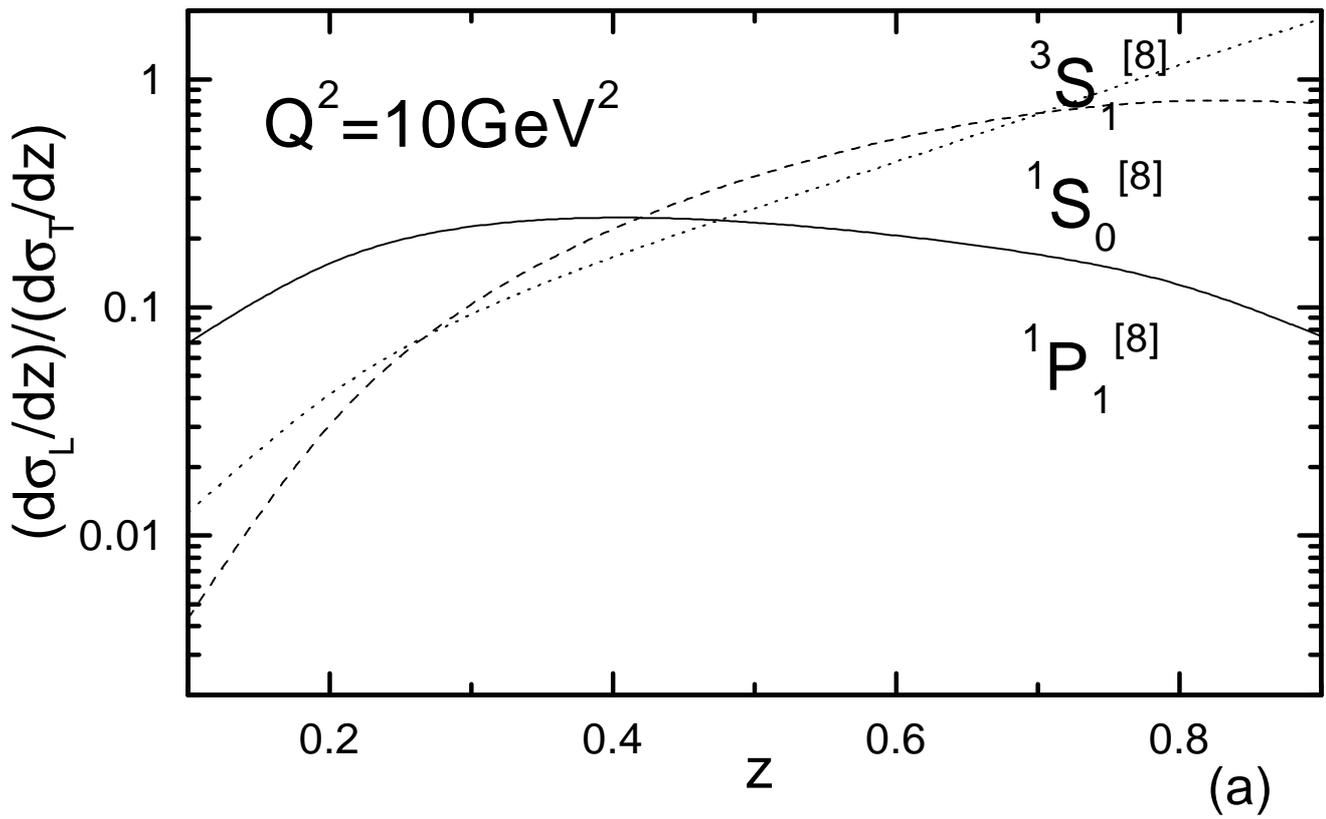
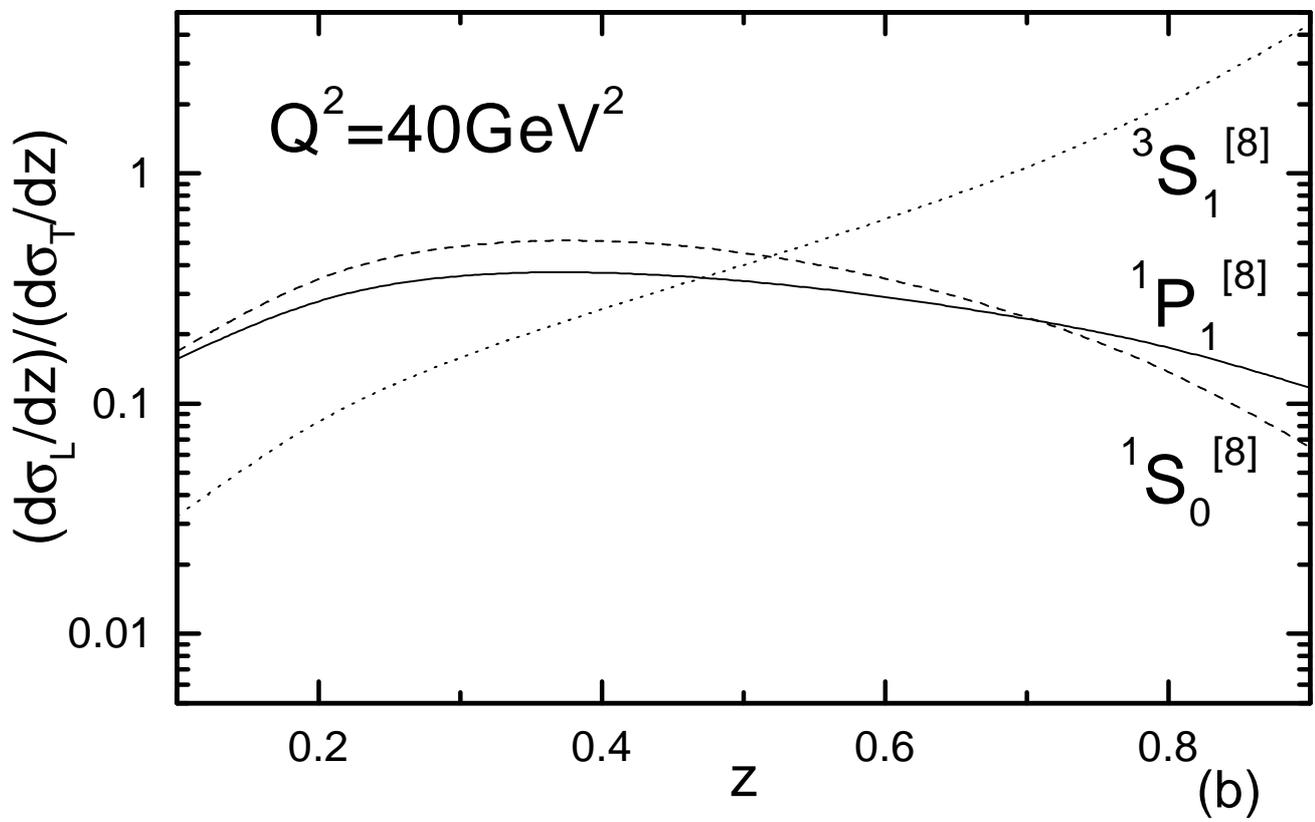
FIG. 4. The ratio $\sigma_L(\gamma^*p)/\sigma_T(\gamma^*p)$ as the function of z. (a) $Q^2=10GeV^2$. (b) $Q^2=40GeV^2$.

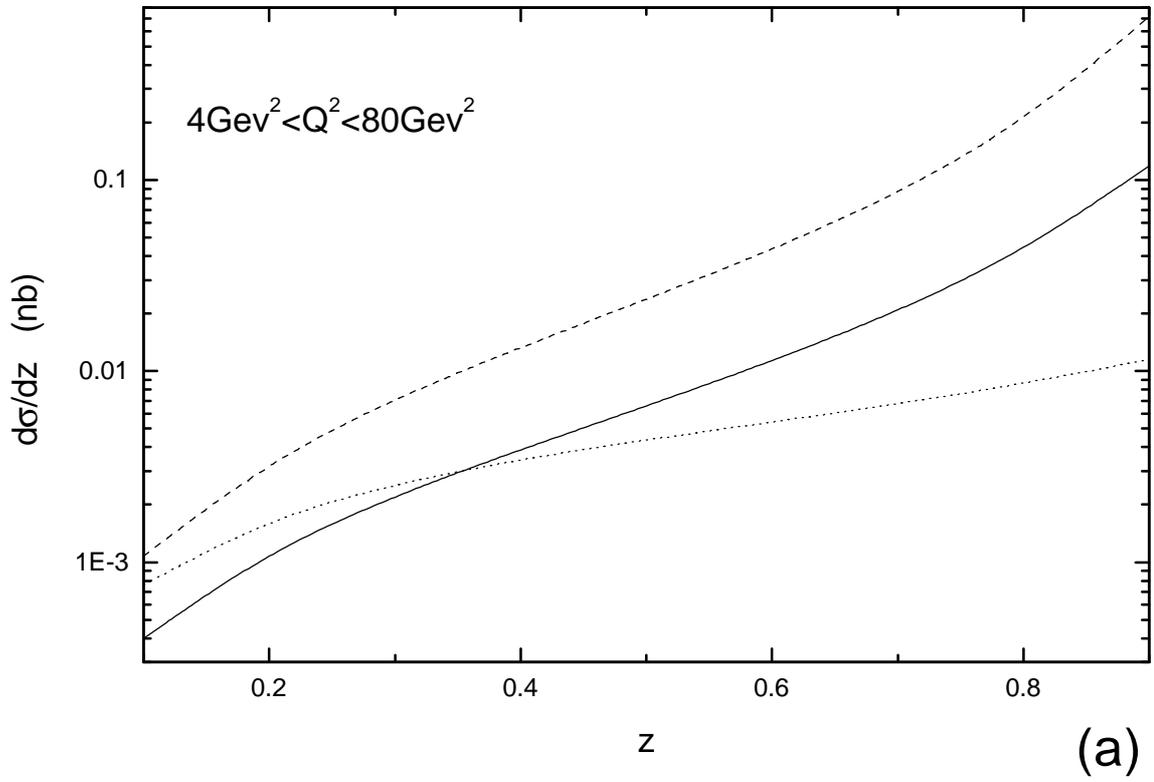
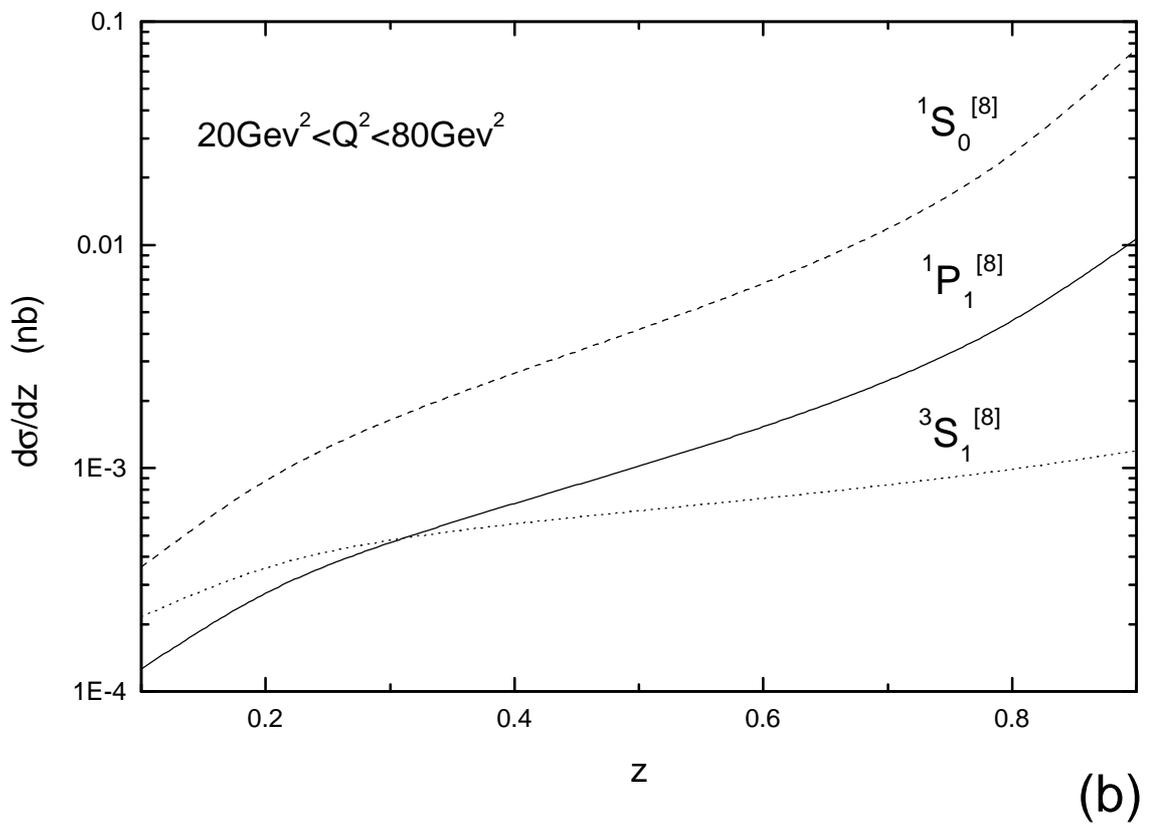
FIG. 5. The cross section $d\sigma /dz$ for $e+p\rightarrow e+\eta_c+X$ at
HERA as a fuction of $z$, where $\sqrt{s_{ep}}=300GeV$.
(a) $4GeV^2<Q^2<80GeV^2$, (b) $20GeV^2<Q^2<80GeV^2$.

\end{document}